\documentclass[12pt,letter]{article}
\usepackage{xcolor}

\usepackage{graphicx}
\usepackage{amsmath}
\usepackage{amssymb}
\usepackage{authblk}
\usepackage[left=1in,right=1in,top=0.8in,bottom=0.8in]{geometry}
\usepackage{setspace}
\usepackage{float}
\usepackage[super,sort&compress,comma]{natbib}
\usepackage[version=3]{mhchem} 
\usepackage{graphicx} 
\usepackage[font={footnotesize}]{caption} 
\usepackage[labelfont=bf]{caption}
\usepackage{amsmath, amssymb}
\usepackage{natbib}
\usepackage{booktabs}
\usepackage{array}
\usepackage{subcaption}
\usepackage{multirow}
\usepackage{soul}
\usepackage{float}
\usepackage{siunitx}
\usepackage{tabularx}
\usepackage{ulem}
\usepackage{hyperref}
\usepackage{xr-hyper}

\begin{document}

\title
{\textbf{Can Charge Transfer Across C–H···O Hydrogen
Bonds Stabilize Oil Droplets in Water?}}

\author{Ruoqi Zhao$^{1,4}$, Hengyuan Shen$^{1}$, R. Allen LaCour$^{1,4}$, Joseph P. Heindel$^{1,4}$, \\
Martin Head-Gordon$^{1,4}$, Teresa Head-Gordon$^{1-4}$}
 \date{}
\maketitle
\begin{center}
\vspace{-10mm}
$^1$Kenneth S. Pitzer Theory Center and Department of Chemistry, $^2$Departments of Bioengineering and $^3$Chemical and Biomolecular Engineering, University of California, Berkeley, Berkeley, CA, 94720 USA\\ 

$^4$Chemical Sciences Division, Lawrence Berkeley National Laboratory, Berkeley, CA, 94720 USA

corresponding author: thg@berkeley.edu
\end{center}

\begin{abstract}
\noindent
Oil-water emulsions resist aggregation due to the presence of negative charges at their surface that leads to mutual repulsion between droplets, but the molecular origin of oil charge is currently under debate. While much evidence has suggested that ionic species must accumulate at the interface, an alternative perspective attributes the negative charge on the oil droplet to charge transfer of electron density from water to oil molecules. While the charge transfer mechanism is consistent with the correct sign of oil charge, it is just as important to provide good estimates of the charge magnitude to explain emulsion stability and electrophoresis experiments. Here we show using energy decomposition analysis that the amount of net flow of charge from water to oil is negligibly small due to nearly equal forward and backward charge transfer through weak oil-water interactions, such that oil droplets would be unstable and coalesce, contrary to experiment. The lack of charge transfer also explains why vibrational sum frequency scattering reports a blue shift in the oil C-H frequency when forming emulsions with water, which arises from Pauli repulsion due to localized confinement at the interface. Finally, unlike ions, neither charge transfer nor dynamic polarization can produce a finite conductivity needed to couple to electric fields that would explain electrophoretic mobility.

\end{abstract}


\begin{figure}[H]
    \centering
    \includegraphics[width=0.45\linewidth]{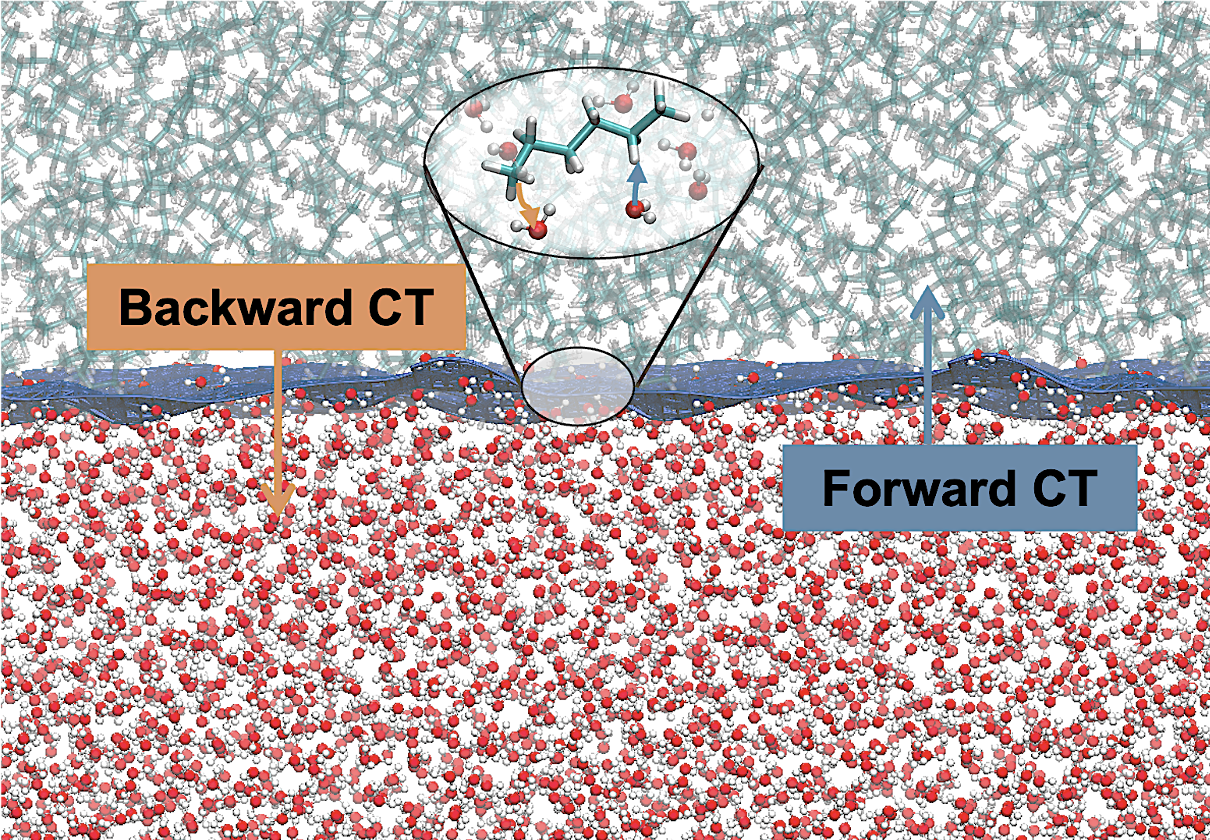}
    \label{fig:interface-snapshot}
\end{figure}

\section{Introduction}
Oil and water are capable of forming long-lived metastable emulsions in which the minor component forms droplets dispersed throughout the solvent. Oil-water emulsions are thought to be electrostatically stabilized by surface charges that lead to mutual repulsion between droplets to confer metastability. For the case of oil droplets in water, electrophoresis experiments consistently show that the oil droplets migrate toward the anode under an applied electric field\cite{carruthers1938,hunter2013zeta}, indicating that the oil droplets preferentially adsorb negative ionic species present in the aqueous solution. This gives rise to a stationary layer of ions that adhere to the oil particle and travel with the droplet under the applied field, creating an electrostatic potential difference with the mobile fluid at the ``slip plane'' which defines the $\zeta$ potential. The $\zeta$ potential can be inferred from the electrophoretic mobility of the droplet and is frequently used as a gauge of emulsion stability\cite{anderson1989colloid,ohshima1994simple}, with $\zeta$ potential magnitudes less than $\sim$30 mV resulting in flocculation or coalescence\cite{kumar2017methods}. For oil-in-water emulsions prepared at neutral pH, the $\zeta$ potential is typically around $-$60 to $-$90 mV,\cite{Vacha2011,agmon2016protons,Shi2025} indicating excellent longer term droplet metastability. 

Because stable oil-in-water emulsions can be prepared without added ions, it is frequently assumed that hydroxide ions from the auto-dissociation of water preferentially adsorb to the water oil-interface.\cite{beattie2009surface,Gan2017} Indeed, the $\zeta$ potential varies strongly with the pH, including going through an isoelectric point around pH = 3.\cite{carruthers1938} In 2004, Beattie and Djerdjev estimated the surface charge density of interfacial ions by combining electroacoustic droplet sizing with measurements of the NaOH required to maintain a constant pH during homogenization.\cite{Beattie2004} They observed that the amount of OH$^-$ added increased linearly with increasing interfacial area, suggesting an intrinsic and constant surface charge density at the oil-water interface. Under the assumption of excess OH$^-$ ions they estimated the surface charge to be $-$0.28 e$^-$/nm$^2$. More recently using second harmonic generation (SHG) experiments, Gan et al. found that hydroxide, not hydronium, showed preferential absorption at the surface of an hexadecane emulsion in water.\cite{Gan2017} Their estimates of OH$^-$ surface charge of $-$0.14 e$^-$/nm$^2$, which is comparable to the value obtained previously by Beattie and Djerdjev, would support an adsorption free energy of $\sim -$8.3 kcal/mol.\cite{Gan2017} Alternatively, some studies have suggested that the interfacial ions originate from trace amounts of surface-active impurities, such as amphiphilic molecules with carboxylic acid groups, which adsorb at hydrophobic interfaces and account for the negative zeta potential and its pH dependence.\cite{Roger2012a} Netz and co-workers further demonstrated that including nanomolar concentrations of weakly acidic and basic impurities could reproduce experimental zeta potentials in theoretical models.\cite{uematsu2019impurity, uematsu2020nanomolar}

However a different perspective has been offered over the last 15 years that attributes the negative charge on the oil droplet not to ionic species, but an accumulation of excess electron density at the interface arising from charge transfer (CT).\cite{Vacha2011} In 2011 Vacha and co-workers\cite{Vacha2011} combined mobility measurements, sum frequency scattering (SFS) experiments, and molecular dynamics simulations. The mobility measurements measured a $\zeta$ potential of $-$81 mV, while the SFS experiments showed no selective adsorption of OH$^-$ ions to the oil-water interface. The molecular dynamics (MD) simulations found that a small amount of charge was transferred between water hydrogen bonds, not between water and oil, using the TIP4P-FQ water model that is capable of modeling CT.\cite{lee2011effects} They calculated a water surface charge density of $-0.0001$ to $-0.0028$ e$^-$/nm$^2$, and subsequent studies employing similar charge equilibration models\cite{wick2012intermolecular,vacha2012charge} also reported comparable surface charge densities in the range of $-0.001$ to $-0.002$ e$^-$/nm$^2$. However, the surface charge estimates from these studies primarily refer to CT interactions between water molecules rather than CT between water and oil, and furthermore are too small in magnitude to support metastability.\cite{Vacha2011,wick2012intermolecular,vacha2012charge}. 

In subsequent work Pullanchery and co-workers introduced vibrational sum frequency scattering (VSFS) measurements on oil-water emulsions\cite{pullanchery2021charge}, and found spectroscopic signatures in which the oil C–H modes blue shift while the water vibrations red shift at the interface. This spectroscopic signature was claimed to be supportive for a CT mechanism, since they again found no experimental evidence from VSFS of interfacial OH radicals, OH$^-$ ions, or contaminant species.\cite{pullanchery2021charge} Supporting theoretical calculations further proposed that CT occurs across C–H···O hydrogen bonds\cite{pullanchery2021charge}. The CT mechanism has also been analyzed from theoretical calculations using the density-derived electrostatic and chemical (DDEC)\cite{manz2012improved} atomic charge partitioning scheme, which found a charge of $-$0.015 e$^-$/nm$^2$, but only after accumulation into the droplet and deep in the oil phase.\cite{poli2020charge} This is an order of magnitude higher than previous charge equilibration calculations that estimated CT surface density .\cite{Vacha2011,wick2012intermolecular,vacha2012charge}  More recently in 2024, Roke and co-workers found that while oil mobility changes as a function of increasing pH, the blue-shifted vibrational signatures of the C–H peak remain unperturbed by increasing pH which they claim confirms a CT mechanism\cite{Pullanchery2024}. To explain the mobility results, they conducted ab initio molecular dynamics (AIMD) simulations and found that under a 0.1 V/~\text{\AA} electric field oriented along the z-axis, a single neopentane molecule exhibited a net drift relative to the surrounding water which they attributed not solely to CT but also to field-induced 'dynamic polarization', to help explain the sensitivity of mobility, but not vibrations, to pH.\cite{Pullanchery2024} 

While the CT offers an intriguing explanation for interfacial charging of the oil phase, there are several reasons to doubt its viability for explaining  stability, electrokinetic experiments, and spectroscopic measurements of oil-water emulsions. To address these concerns, we employ model cluster calculations as well as a polarizable force field simulation of an extended water–hexane interface to quantify the amount of CT using quantum mechanical energy decomposition analysis based on the absolutely localized molecular orbital (ALMO-EDA) method\cite{khaliullin2008analysis,horn2016probing,Mao2018,Mao2019,Mao2021}. Our findings indicate that the surface charge densities computed by ALMO-EDA are one to two orders of magnitude smaller than the DDEC atomic charge partitioning scheme or charges obtained from Mulliken populations. Furthermore, true CT analysis shows any CT from water to oil is largely negated by back-transfer of electron density\cite{loipersberger2020variational} from oil to water, in which back-transfer is a natural energy lowering quantum mechanical process when one correctly accounts for electron flow. Therefore such small amounts of net CT can't possibly explain oil-water emulsion stability and hence does not explain oil charging and electrophoresis experiments. We reaffirm again that vibrational blue shifts of C-H actually arise from Pauli repulsion whereas CT would redshift these same modes\cite{gu1999fundamental,li2002physical,Mao2019}, and thus a CT mechanism can't be the origin of the VSFS spectroscopic observations. Finally we show that explanations based on dynamic polarization to explain mobility\cite{Pullanchery2024} lack evidence due to AIMD simulation errors that fail to exhibit directional drift under an applied electric field. This is important for the reason that neither the small density from CT nor the polarization fluctuations can sustain the conductivity needed to couple into the laterally applied electric fields to explain electrophoresis experiments.
\vspace{-3mm}

\section{Results}
Beginning with accurate state-of-the-art density functional theory (DFT) calculations\cite{Mardirossian2017}, the ALMO-EDA framework\cite{khaliullin2008analysis,horn2016probing,Mao2018,Mao2019,Mao2021} decomposes the overall intermolecular interaction energy, $\Delta E_\text{INT}$ between several species or fragments into three physically interpretable contributions, providing insight into their relative significance. 

\begin{equation}
    \Delta E_\text{INT}=\Delta E_\text{FRZ}+\Delta E_\text{POL}+\Delta E_\text{CT}
\end{equation}

\noindent
The frozen interaction term ($\Delta E_\text{FRZ}$) is comprised of Pauli repulsion, dispersion, and permanent electrostatics. We also separate polarization ($\Delta E_\text{POL}$) from CT, as they have different distance dependence and physical origins. $\Delta E_\text{POL}$ is the internal charge rearrangements or charge shifts of individual fragment in the presence of the others without charge flow between them. The final self consistent field (SCF) DFT calculation without any constraint enables electron density exchange between fragments and allows for the determination of forward CT from water$\to$oil, as well as backward CT from oil$\to$water.

Within the ALMO-EDA framework, electron density rearrangement associated with CT can be interpreted as donor-acceptor orbital interactions. CT occurs when an occupied orbital on one molecule donates electron density to a complementary virtual orbital on another molecule. This interaction is quantified through complementary occupied-virtual pair (COVP)\cite{khaliullin2008analysis,veccham2021non,shen2022generalization} analysis. In what follows we consider a single dominant forward COVP and a single dominant backward COVP, because they are sufficient to capture approximately 75\% of the overall CT contribution to both the energy and net charge, making them a compact summary of the CT orbital picture. A more detailed description of the ALMO-EDA method, as implemented in Q-Chem\cite{Epifanovsky2021}, and further details of forward and backward CT using COVP analysis are provided in the Methods section.

\subsection{Charge transfer across the C-H···O hydrogen bond}
\begin{figure}[H]
    \centering
    \includegraphics[width=1.0\linewidth]{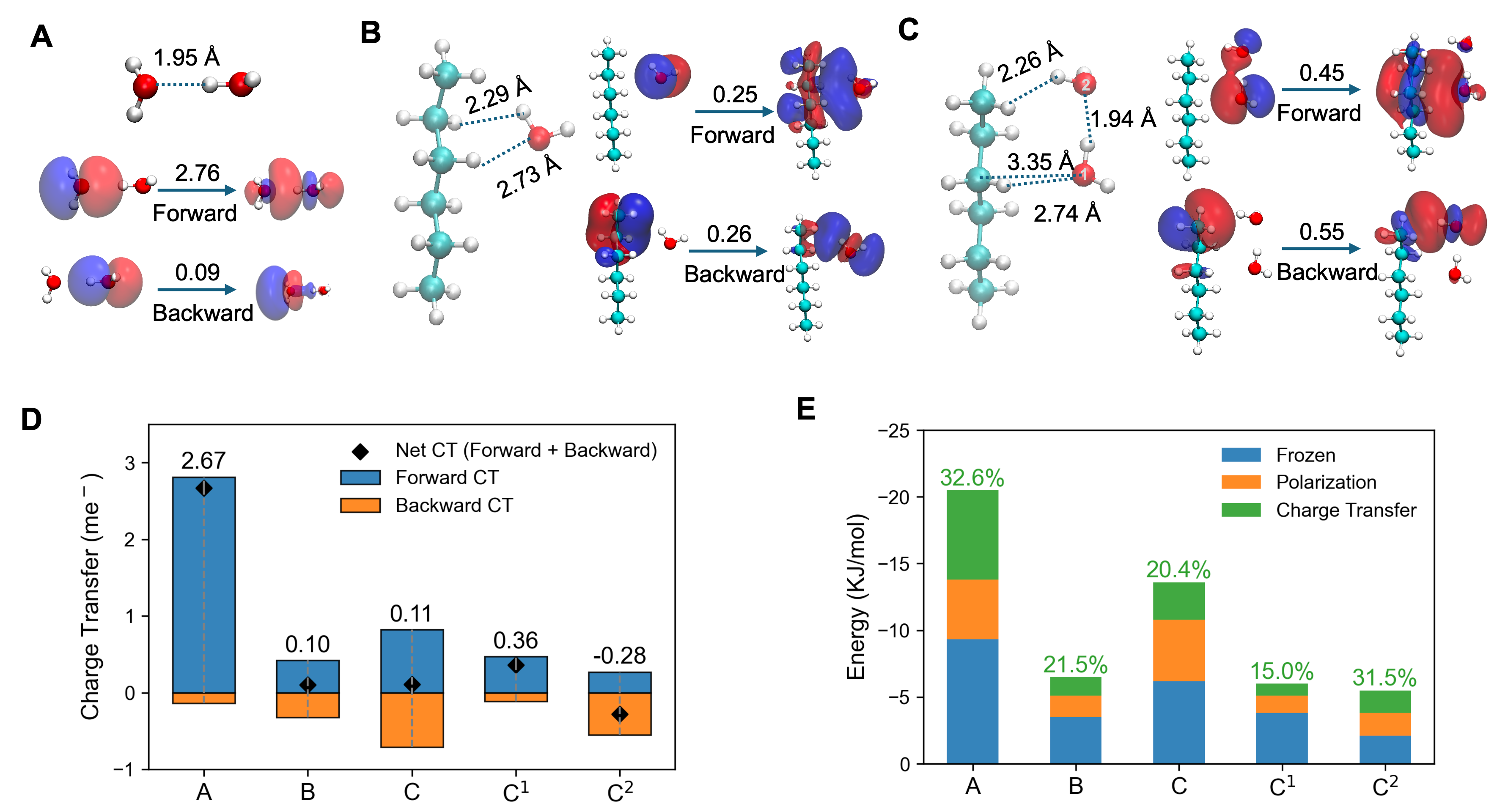}
    \caption{\textit{Charge transfer analysis for optimized water–hexane complexes compared to the  water dimer}. Optimized structures and dominant COVPs in forward and backward charge transfer processes of \textbf{A} (H$_2$O)(H$_2$O), \textbf{B} (C$_6$H$_{14}$)(H$_2$O), and \textbf{C}(C$_6$H$_{14}$)(H$_2$O$^1$---H$_2$O$^2$) complexes. Forward CT refers to H$_2$O $\rightarrow$ C$_6$H$_{14}$, and backward CT refers to C$_6$H$_{14}$ $\rightarrow$ H$_2$O. \textbf{D} and \textbf{E} represent charge transfer and ALMO-EDA energy analysis of optimized complexes in \textbf{A}, \textbf{B}, \textbf{C}, and \textbf{C$^1$} and \textbf{C$^2$} that represent the complexes (C$_6$H$_{14}$)(H$_2$O$^1$) and (C$_6$H$_{14}$)(H$_2$O$^2$) after removing one water molecule from complex \textbf{C}. All structural optimizations were performed at the $\omega$B97X-V/def2-TZVP level, and AMLO-EDA calculations were conducted using $\omega$B97X-V/def2-TZVPD. CT unit me$^-$. The original data for \textbf{D} and \textbf{E} are in Supplementary Tables S1 and S2.}
    \label{fig:small-cluster}
\end{figure}

\vspace{-2mm}
We begin with the prototype system for understanding CT based on the strong hydrogen bond O-H···O formed by the the water dimer, in order to compare it to C-H···O interactions in which it is claimed that CT occurs from water to oil\cite{pullanchery2021charge}. As seen in Figure \ref{fig:small-cluster}A, the COVP indicates that hydrogen bonding in the water dimer is composed of CT from an electron pair localized on an oxygen atom donating electrons to the O–H $\sigma$ anti-bonding orbital on the other water. For the water dimer, CT occurs predominantly in the forward direction, from the hydrogen-bond acceptor water molecule to the donor water molecule, with a net CT of 2.67 me$^-$ quantified in full by ALMO-EDA. 

To accurately assess CT in oil–water interactions, we next analyze the C-H···O interaction that has been widely recognized as another, albeit weaker, type of  hydrogen bond.\cite{gu1999fundamental,horowitz2012carbon} Many studies impose a linear constraint when analyzing C-H···O interactions\cite{gu1999fundamental}, however their relatively weak strength makes them more susceptible to bending and less sensitive to deviations from their equilibrium bond length\cite{gu1999fundamental}. This is seen in the optimized structures of a hexane molecular fragment (C$_6$H$_{14}$) interacting with one water molecule as shown in Figure \ref{fig:small-cluster}B, in which the optimized structure exhibits a C-H···O bond angle of approximately 120°, with the hydrogens in water positioned closer to hexane than the oxygen atoms. This significantly deviates from the linear hydrogen bond configuration observed in the water dimer or idealized C-H···O configurations. 

The accompanying dominant COVPs from ALMO-EDA show that forward CT (0.25 me$^-$) from the $\pi$ type lone pair electrons of water's oxygen is an order of magnitude smaller than for the water-water hydrogen bond because the C-H group is a poor donor for hydrogen-bonding. 
Moreover, a backward CT of 0.26 me$^-$ from hexane to the $\sigma^\star$ orbital of water nearly compensates for the forward CT. Given the close spatial proximity of water hydrogen atoms to hexane, electron density flow from the oxygen of water toward hexane while simultaneously being drawn back from hexane to the hydrogen atom of water, results in a negligible 0.01 me$^-$ transfer of charge. Even if we are to include all forward and backward COVP components, Figure \ref{fig:small-cluster}D indicates that the total net CT only amounts to 0.10 me$^-$, i.e, 25 times smaller than the water dimer. Consequently, it is not surprising that the optimized hexane-water structure does not exhibit the typical linear hydrogen-bonding configuration of the water dimer that is strengthened by a large net CT interaction. 

\vspace{-2mm}
In Figure \ref{fig:small-cluster}C we see that the presence of a second water molecule enhances the forward CT by a factor of $\sim$1.8  compared to the hexane-water dimer in Figure~\ref{fig:small-cluster}B. This enhancement arises from the formation of a stronger electron donor through hydrogen bonding with the second water molecule. However, the increased forward CT is accompanied by an increase in backward CT, as the second water molecule becomes more electron-deficient, with the virtual orbitals in the backward CT primarily localized on the second water molecule. In this case the C-H group returns electron density to the anti-bonding orbital of the second water given the close range of H··H interactions. To further investigate the role of the C-H···O hydrogen bond geometry in the CT mechanism, we remove one water molecule from the (C$_6$H$_{14}$) (H$_2$O$^1$---H$_2$O$^2$) to create (C$_6$H$_{14}$)(H$_2$O$^1$) (\textbf{C$^1$}) and removing the other water molecule to create (C$_6$H$_{14}$)(H$_2$O$^2$) (\textbf{C$^2$}). As seen in Figure \ref{fig:small-cluster}C, \textbf{C$^1$} has a well-defined hydrogen bond with a C··O distance of $\sim$3.35\AA, resulting in a net CT of 0.36 me$^-$, whereas \textbf{C$^2$} lacks an optimal hydrogen bond configuration, leading to a net CT in the reverse direction from hexane to water of $-$0.28 me$^-$, which explains the negligible net charge flow in Figure \ref{fig:small-cluster}C. These findings indicate that even though the presence of an additional water molecule enhances forward CT, the overall net electron transfer remains negligibly small, and furthermore it accumulates on water not hexane. 

Finally it is important to emphasize that both forward and backward CT processes stabilize the system and contribute substantially to lowering its energy. As shown in Figure \ref{fig:small-cluster}E, the CT interaction energy plays a similarly crucial role in stabilizing the two water–hexane complex as it does in the water dimer. But unlike the water dimer, forward and backward CT density are nearly equal in magnitude but opposite in sign, resulting in a mostly balanced charge exchange between water and hexane. This challenges the assumption that the presence of a C-H···O hydrogen bond necessarily induces significant, directional charge transfer from water to oil. 

\subsection{Charge transfer at an extended oil-water interface}
\vspace{-2mm}
Analysis of minimized dimer and trimer configurations reveal that forward and backward CT between water and hexane nearly cancels. However, the CT mechanism at the extended water–oil interface may be significantly more intricate due to thermal fluctuations and a more extended network of hydrogen-bond interactions between waters and hydrophobic association of hexanes. To gain further insight into the CT process at an extended water–oil interface, molecular dynamics at ambient conditions using the polarizable AMOEBA force field\cite{ponder2010current,ren2011polarizable} was employed; simulation details are provided in the Methods section. 

\begin{figure}[H]
    \centering
   \includegraphics[width=0.975\linewidth]{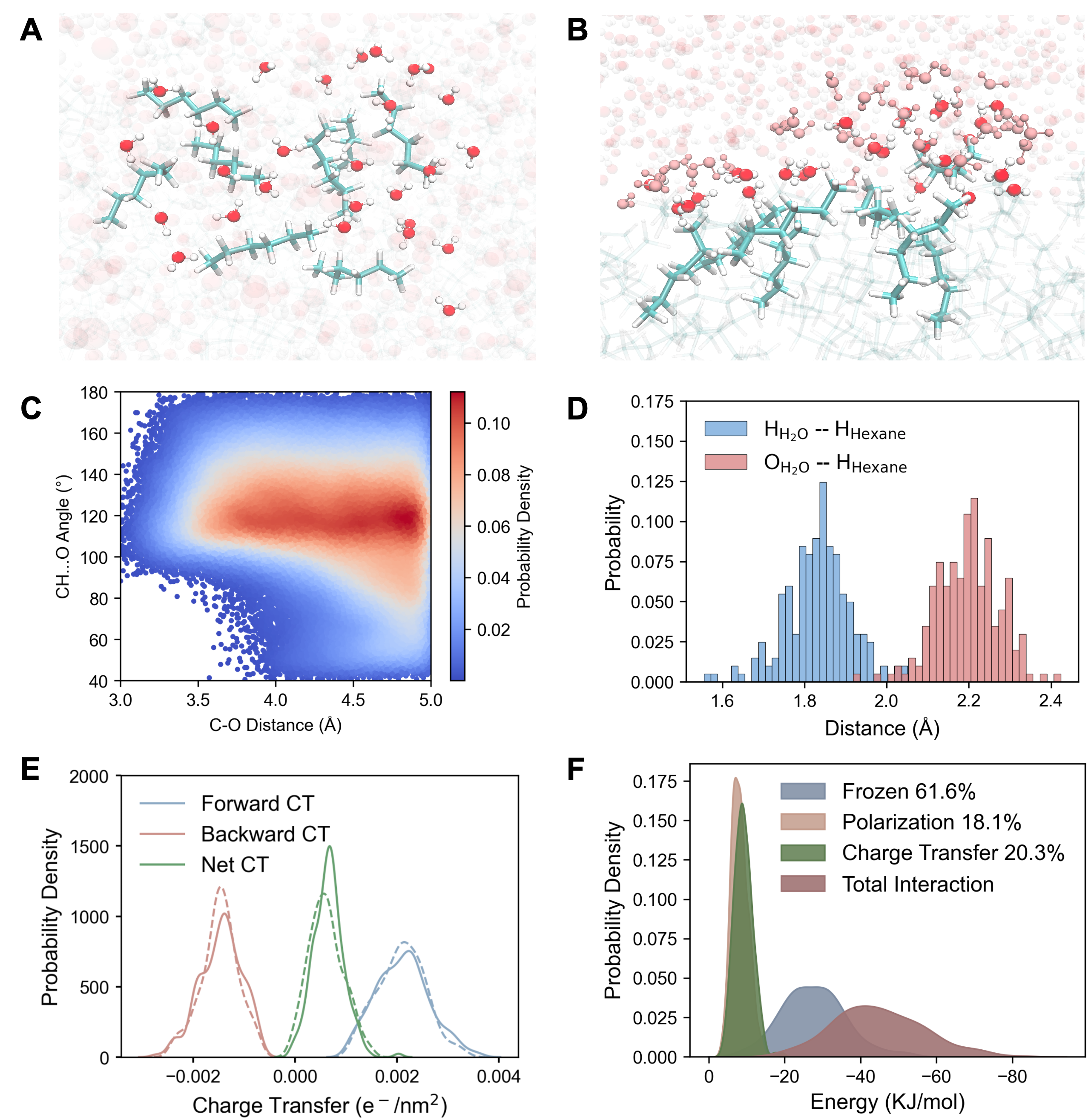}
    \caption{ $\textbf{Water/oil interface structures}$. ($\textbf{A, B}$) Sample representation of the clusters considered for the ALMO-EDA calculations highlighting the spatial arrangement of water molecules (red and white spheres) around hexane molecules (cyan). The extended hydrogen-bonded water network at the interface is illustrated in pink. ($\textbf{C}$)  Probability density distribution illustrating the spatial and angular relationship between water oxygen and hexane. The x-axis represents the C–O distance, while the y-axis shows the C-H···O angle, describing the relative orientation of water molecules with respect to hexane’s C–H bonds. ($\textbf{D}$) Probability distribution of the closest distance between hydrogen atoms in hexane with water hydrogens and oxygens. This is consistent with the radial distribution functions (Supplementary Figure S1), where the H···H intermolecular distribution is more short-ranged than the O···H distribution in the water–hexane system. ($\textbf{E}$) Probability density distributions of Forward CT, Backward CT, and Net CT. Solid lines for direct water–hexane interactions clusters in (A) and dashed lines correspond to the clusters in (B). The original data is provided in Supplementary Tables S3 and S4. ($\textbf{F}$) Probability density distributions of water oil inter-molecule interaction energy components. The legend includes the percentage contribution of each energy component relative to the total interaction energy. Forward CT refers to H$_2$O $\to$ C$_6$H$_{14}$, and Backward CT refers to C$_6$H$_{14}$ $\to$ H$_2$O. Cluster selection details are provided in the Methods section. }
    \label{fig:interface-structure}
\end{figure}

From the MD simulations we draw two types of interfacial hexane-water clusters. Figure \ref{fig:interface-structure}A is a representative snapshot of the 200 clusters in which the water molecules are localized around multiple hexanes, whereas Figure \ref{fig:interface-structure}B is a representative snapshot of the 50 clusters in which interfacial water molecules form more extensive hydrogen-bonded network around the hexane. From these clusters we derive the probability density distribution in Figure \ref{fig:interface-structure}C that shows a high-probability region near a CH...O angle of 120° for C-O distances $>$ 3.5 \AA, consistent with the C-H···O configurations of the ab initio optimized clusters in Figure \ref{fig:small-cluster}B,C. The broad distribution means that the C-H···O hydrogen bond is significantly weaker and more distorted compared to the highly directional hydrogen bonds found for thermalized water–water interactions which experience significant CT. This would indicate that CT in the forward direction from water to oil will be much smaller at an extended oil-water interface as well. In Figure \ref{fig:interface-structure}D, the distribution of distances between hexane and water hydrogens is centered at 1.85~\text{\AA}, whereas the distribution of water oxygen to hexane hydrogen distances is peaked at 2.2~\text{\AA}, and the two distributions exhibit minimal overlap. This interfacial organization implies that the hydrogen of surface water orientated towards the oil (``dangling O-H groups'') will be a conduit for back-transfer of electron density from oil to water.

This is borne out in Figure \ref{fig:interface-structure}E which illustrates the ensemble nature of CT dynamics at the oil-water interface. The forward CT distribution of the interfacial clusters peaks at 2.09 me$^{-}$/nm$^{2}$, which arises from the C-H···O hydrogen bond at the interface, and similar to the small clusters in Figure \ref{fig:small-cluster}. But again this is balanced by electron backflow from the CT distribution centered around -1.44 me$^{-}$/nm$^{2}$ due to the presence of short-range H···H interactions at the interface that defines the source of backward CT as also seen in Figure \ref{fig:small-cluster} The average net CT of 0.00065 e$^{-}$/nm$^{2}$ (0.65 me$^{-}$/nm$^{2}$) obtained from ALMO-EDA corresponds to 0.00007 e$^{-}$/hydrogen bond, closely matching the 0.0001 e$^{-}$/hydrogen bond 

\noindent
observed in the hexane–water clusters in Figure \ref{fig:small-cluster}. Furthermore these conclusions are consistent between the clusters with localized waters near hexane or the clusters incorporating the extended hydrogen bonding network since interfacial CT is predominantly dictated by direct water–hexane interactions, whereas the broader hydrogen bonding network within the interfacial region exerts only a subtle perturbative influence (average net CT of 0.00061 e$^{-}$/nm$^{2}$). Finally Figure \ref{fig:interface-structure}F shows that the frozen energy term constitutes the dominant component (61.6\%) to water-hexane interactions, with similar percentages of polarization energy and CT making up the remaining energetic interactions. The estimated 20.3\% of CT energy between water and hexane is a proportion similar to that observed in Figure \ref{fig:small-cluster} for small water–hexane clusters B and C, and is energetically the manifestation of electron delocalization in stabilizing the system. 

\subsection{Surface charge and oil-water emulsion metastability}
\vspace{-2mm}
Using the non-linear Poisson Boltzmann equation\cite{Gouy1910,Chapman1913,Debye1923,ohshima1982accurate}, we consider how different estimates of surface charge give rise to changes in surface potential which is a measure of emulsion stability. For ionic species it is appropriate to use the non-linear Poisson Boltzmann equation\cite{Gouy1910,Chapman1913,Debye1923,ohshima1982accurate}, which connects the surface charge density \( \sigma_0 \) to the surface potential \( \psi_0 \):
\begin{equation}
\sigma_0 = Z \sqrt{2 \varepsilon_r \varepsilon_0 k_B T \rho_\text{bulk}} 
\left[
2 \sinh\left( \frac{e \psi_0}{2 k_B T} \right)
+ \frac{4}{R \left( \frac{\varepsilon_0 \varepsilon_r k_B T}{2 e^2 Z^2 \rho_\text{bulk}} \right)^{-1/2}}
\tanh\left( \frac{e \psi_0}{4 k_B T} \right)
\right],
\end{equation}
where \( \varepsilon_r \) is the static dielectric constant of bulk water, \( k_B \) is the Boltzmann constant, \( T \) is the temperature,  \( e \) is the elementary charge, and \( R \) is the droplet radius. We note that previous experiments have reported average surface droplet radii of 100 nm\cite{pullanchery2021charge,Shi2025} to 230 nm\cite{pullanchery2020stability,Pullanchery2024}; here we use a radius of 125 nm as reported most recently\cite{Pullanchery2024,Shi2025}. We also consider the relationship between surface charge and surface potential under different ionic concentrations, where Z is the ion valence and \( \rho_\text{bulk} \) is the bulk ion density. While a concentration of OH$^-$ ions of 0.0001 mM/L is the ionic strength of pure water, surface potential and electrokinetic measurements are also conducted in the presence of electrolytes and hence at higher ionic strengths that also must be reconciled for any mechanism. 

\begin{figure}[H]
    \centering
    \includegraphics[width=0.99\linewidth]{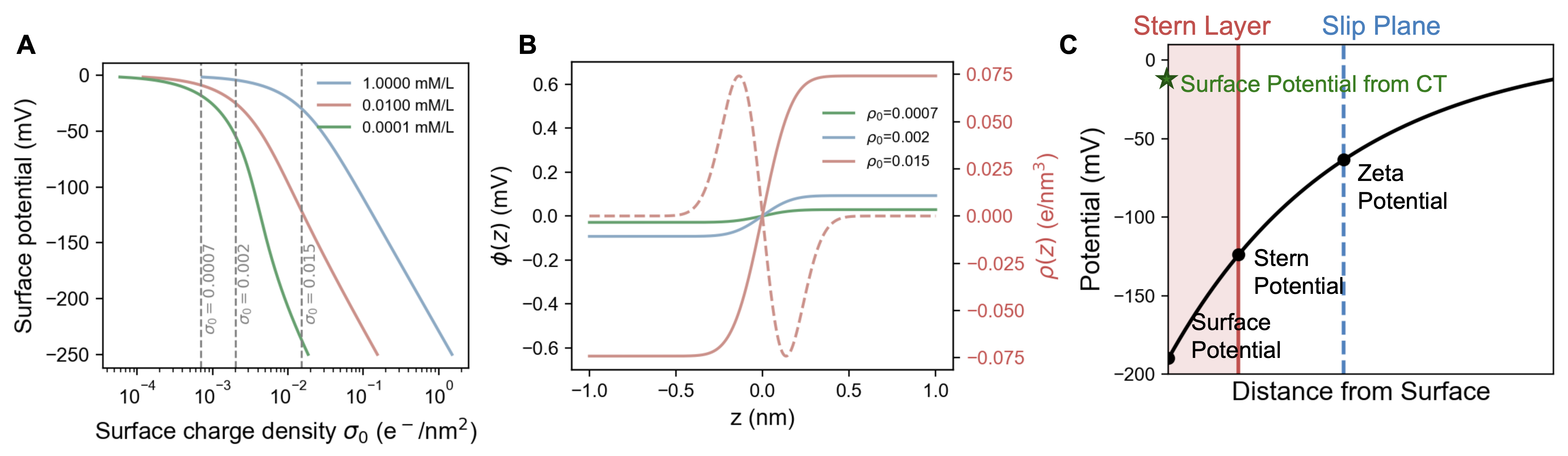}
    \caption{ $\textbf{Estimates of surface charge and zeta potentials}$. ($\textbf{A}$) Surface potential as a function of surface charge density ($\sigma_0$) for a 100 nm-radius spherical interface at various ionic strengths (1 mM, 0.01 mM, 0.0001 mM) using the non-linear Poisson Boltzmann equation\cite{Gouy1910,Chapman1913,Debye1923}. Dashed vertical lines are surface charge densities of $-$0.015 \cite{poli2020charge}, $-$0.002 \cite{vacha2012charge}, and $-$0.0007 e$^-$/nm$^2$ from ALMO-EDA. See Supplementary Information.($\textbf{B}$) The surface potential (solid lines) as a function of surface charge density using the Poisson equation.}  ($\textbf{C}$). An illustration of the potential at the surface as assumed by a CT mechanism, and due to the double layer or beyond under the assumption of ionic species such as hydroxide ions. Adapted with permission from\cite{lacour2025role}. Copyright 2025 American Chemical Society.
    \label{fig:zeta}
\end{figure}

Figure \ref{fig:zeta}A shows that the surface charge density of -0.0007 e$^-$/nm$^2$ found from ALMO-EDA, which defines a true quantum mechanical definition of CT, would predict an unstable emulsion at any value of ionic strength, as the surface potential would be below $-$30 mV. Even under the assumption of a high surface charge density of $-$0.015 e$^-$/nm$^2$ suggested by Poli et al.,\cite{poli2020charge} the emulsions would be unstable at moderately high pH or salt strength, also in contradiction to experiments (i.e. an oil droplet in 1mM NaOH or NaCl bulk solution as reported\cite{Pullanchery2024}). Oil-water emulsions are in fact metastable over weeks to months\cite{Shi2025} at at all stated pH values, so surface charge is required to be larger as often advocated by Beattie and co-workers.\cite{beattie2009surface} 

But because CT is short-ranged, the surface charge density is a very thin double layer such that the macroscopic Poisson-Boltzmann model is not warranted. Instead we use the integral form of the one-dimensional Poisson equation\cite{becker2022} 
\begin{equation}
\psi(z) = -\frac{1}{\varepsilon_0 \varepsilon_r} 
\int_{z_1}^z \left[ \int_{z_1}^{z'} \rho_e(z'') \, dz'' \right] dz'
\end{equation}
to integrate over the Gaussian distributed charge over the narrow interface to describe the double layer under the CT assumption.The Gaussian-distributed charge density is generated based on an approximation from the surface charge density, representing the spatial distribution of charges along the surface normal. This is illustrated for the $-$0.015
$e-/nm{^2}$ value in Figure \ref{fig:zeta}B (red dashed line) which is very similar to the reported AIMD simulations by Poli et al\cite{poli2020charge}. As also seen in Figure \ref{fig:zeta}B the resulting surface potential (solid lines) is very small and unconducive to metastability, and more realistic estimates of surface charge from ALMO-EDA simply reinforces this conclusion. Even changing the bulk dielectric of 80 to the high frequency limit of 1.8 under the assumption that dipolar fluctuations are completely damped out at the interface would still result in surface potentials well below $-$30 mV.

Electrophoresis experiments performed on oil-water emulsions measure the electrokinetic mobility of the oil droplet, $\mu_{e}$ and its connection to the $\zeta$ potential to explain emulsion metastability (Figure \ref{fig:zeta}C). The lateral electric field applied in an electrokinetic experiment must couple to an electric current\cite{Bonthuis2010}, for which mobile charges are needed in order to distinguish them from bound charges, and to justify their use in the Helmholtz-Smoluchowski equation
\begin{equation}
\mu_{e} = \varepsilon_0 \varepsilon_r \zeta /\eta
\end{equation}
\noindent
For ionic species, using the bulk water dielectric constant and viscosity in Eq. 4 yields good agreement with electrokinetic experiments for oil-water emulsions. Under the CT assumption, however, it is clear that Eq. 4 is invalid for a number of reasons. First is because it requires the localized CT electron density to be divided into immobile and mobile regions to create a slip plane which undergoes shear to generate mobility.\cite{lacour2025role} Another concern is that at an interface with immobile charges one would have to use a value closer to the high-frequency water dielectric constant in Eq. (4) since water orientational and vibrational contributions have been damped out, and again reducing mobility by a factor of 20 to 40 that is not in line with experiments.

It is also possible that the CT density on the oil is fully bound and negligibly small, and thus does not produce a finite conductivity, and hence there is no $\zeta$ potential or mobility at all. In fact, Pullanchery et al.\cite{Pullanchery2024} also reported an AIMD simulation in which a neopentane molecule in water moved in the presence of an applied electric field. However, we analyzed their uploaded AIMD trajectories from their publication, finding that the observed net drift corresponded to the center-of-mass motion of the entire simulated box, but not neopentane itself. This is because the oil carries no significant net charge from CT, and it thus explains why there is a lack of net drift under an applied electric field in their AIMD simulations (see Supplementary Figure S2 and discussion).\cite{Pullanchery2024} 

Finally, the absence of net drift indicates that the electric-field-induced time-dependent polarization (also referred to as ``dynamic polarization'') of neopentane alone cannot produce sustained directional forces as reported.\cite{Pullanchery2024} Although dynamic polarization may give rise to transient forces when a molecule is exposed to an electric field, these effects are inherently short-lived, whereas the electrophoretic mobility reflects a steady-state phenomenon governed by the linear response of charges within the slip plane that defines the zeta potential. Since no drift is observed in their DFT simulations, consistent with other AIMD simulations of the air-water interface where CT is operative that also found no field-induced flow\cite{becker2022}, the CT mechanism nor dynamic polarization produces a finite conductivity that can give rise to a $\zeta$ potential.

\section{Discussion and Conclusion} 
Quantifying CT is rigorously possible using computational quantum chemistry calculations, which provide direct access to not only the electron density but the wavefunction and a molecular orbital basis to describe quantum mechanical CT. Our results from ALMO-EDA\cite{khaliullin2008analysis,horn2016probing}, defined as electron relaxation from an optimally polarized state to a delocalized state, allows for both forward and backward CT contributions to lowering the system's energy. In the water dimer, CT is predominantly unidirectional, with the hydrogen bond acceptor oxygen transferring charge to the donor hydrogen, while backward CT is almost negligible. In contrast, forward CT in the water–hexane complex occurs from the hydrogen bond acceptor oxygen to the donor hydrogen in hexane, but this is largely balanced by backward CT from hexane to the hydrogen of water. This bidirectional nature of CT weakens the directionality of hydrogen bonding, contributing to the more flexible and distorted hydrogen-bonding geometry observed at the water–oil interface, and most relevant here, yields a negligible surface charge value of $-$0.0007 e$^-$/nm$^2$ that accumulates from oil to water. As we have shown, such insignificant amounts of surface charge would result in an instability of the oil-water emulsion, in contradiction to experimental observations. 

Atomic charge partitioning methods assign electron density to individual atoms, which therefore includes a larger effect of partitioning the charge distribution of the fully polarized state.\cite{khaliullin2008analysis} This leads to an overestimation of the CT value and an unphysical estimate of the effective donor-acceptor energy gap\cite{khaliullin2009electron,khaliullin2013microscopic}. Hence the very large surface charge of $-$0.015 e$^-$/nm$^2$ from charge partitioning schemes used by Hassanali and co-workers are inherently flawed.\cite{poli2020charge,pullanchery2021charge} This is evident from the slow and smooth 1/r decay of charge accumulation deep into the oil phase in their theoretical analysis (see Figure 2\cite{poli2020charge} and Figure 8\cite{Pullanchery2024}), behavior that instead signifies electrostatic screening rather than what is expected for true CT\cite{khaliullin2013microscopic} which is highly localized at the interface as we have shown above. 

Beyond quantifying the magnitude of charge transfer, another question is whether such interfacial electron density can be shear-separated by a lateral electric field to produce steady-state electrokinetic flow. This would require the transferred charge on one molecule to hop onto neighboring molecules, enabling lateral current flow under an applied field.\cite{Bonthuis2010} However, the DFT-based AIMD simulations reported by Poli et al. exhibit no net drift for neopentane in water\cite{poli2020charge}, and is consistent with other AIMD simulations which have shown no field-induced flow.\cite{becker2022} Regarding the pH dependence, it has been suggested that pH modifies electronic polarization of the bulk solution which induces mobility differences between OH$^-$ and Cl$^-$ ions. However, it is known that Br$^-$ and I$^-$ both have a greater polarizability than Cl$^-$ but there is almost no change in the mobility and zeta potential among the ions. A second explanation invokes the Grotthuss mechanism, suggesting that the conductivity of OH$^-$ is higher than that of Cl$^-$. Yet the molar conductivity of OH$^-$ is lower than that of H$_3$O$^+$\cite{rodriguez2015exact}; if mobility scales with conductivity, acidic solutions should exhibit higher mobility compared to OH$^-$. 

Furthermore, previous and unambiguous EDA studies have explicitly stated that CT would red shift, not blue shift, the C-H vibrational modes of hydrophobic groups interacting with water.\cite{gu1999fundamental,li2002physical,Mao2019} Given the experimental observation by Roke and co-workers that their C-H vibrations blue shift, then CT as a mechanism must be ruled out, and instead is attributable to Pauli repulsion that arises as a confinement effect at the interface between oil and water. This is supported by the observation that the "frozen" term in the ALMO-EDA is a majority of the energetic stabilization from C-H···O hydrogen bonds as we have shown in Figures 1 and 2. This is one of the necessary conditions for the VSFS spectroscopic features to be observed to blue-shift, i.e. relatively weak CT that is insufficient to compensate for the blue-shifting effect of the frozen interaction.\cite{Mao2019}

Since the CT mechanism is clearly unviable as an explanation of emulsion stability and electrophoresis, as well as the VSFS measurements themselves, what is the molecular explanation for oil charge? The net oil charge must be large as it has been shown to give rise to large electric fields of $\sim$ $-$55 to $-$90 MV/cm at the interface.\cite{Shi2025} In our view it is still an open question at present as to the molecular origin that enhances the surface affinity of negatively charged species to the oil-water and related hydrophobic interfaces, and we must consider all available evidence regarding the remaining hypotheses of ionic charge from water dissociation\cite{Beattie2004,mccarty2008electrostatic,gray2009explanation, beattie2012,beattie2014a,Gan2017,pullanchery2021charge,Djerdjev2025} or arising from surface impurities\cite{Roger2012,uematsu2019impurity,uematsu2020nanomolar}. Both explanations face serious challenges and yet neither should be dismissed as the explanation at this stage. We hope to provide a clarifying review on this topic in the near future.

\vspace{5mm}

\noindent
\textbf{DATA AVAILABILITY}. All data is available upon request from the authors.
 
\vspace{5mm}

\noindent
\textbf{AUTHOR CONTRIBUTIONS}. R.Z. and T.H.-G. conceived the scientific direction of the project. R.Z. and H.S. performed all calculations. R.Z., H.S., M.H.-G., and T.H.-G performed analyses. R.Z. and T.H.-G. wrote the manuscript, and all authors provided comments on the results and manuscript.

\vspace{5mm}

\noindent
\textbf{ACKNOWLEDGMENT}. We thank the U.S. National Science Foundation through Grant CHE-2313791 for support of the EDA work and analysis. We thank the CPIMS program, Office of Basic Energy Sciences, Chemical Sciences Division of the U.S. Department of Energy under Contract DE-AC02-05CH11231 supporting research on the interfacial oil-water emulsions. This work used computational resources provided by the National Energy Research Scientific Computing Center (NERSC), a U.S. Department of Energy Office of Science User Facility operated under Contract DE-AC02-05CH11231, and the Lawrencium computational cluster resource provided by the IT Division at the Lawrence Berkeley National Laboratory (Supported by the Director, Office of Science, Office of Basic Energy Sciences, of the U.S. Department of Energy under Contract No. DE-AC02-05CH11231).

\bibliography{reference}
\bibliographystyle{naturemag}

\end{document}


\title
{\textbf{Supplementary Information: Can Charge Transfer Across C–H···O Hydrogen
Bonds Stabilize Oil Droplets in Water?}}

\author{Ruoqi Zhao$^{1,4}$, Hengyuan Shen$^{1}$, R. Allen LaCour$^{1,4}$,Joseph P. Heindel$^{1,4}$, Martin Head-Gordon$^{1,4}$, Teresa Head-Gordon$^{1-4}$}
 \date{}
\maketitle
\begin{center}
\vspace{-10mm}
$^1$Kenneth S. Pitzer Theory Center and Department of Chemistry, $^2$Departments of Bioengineering and $^3$Chemical and Biomolecular Engineering, University of California, Berkeley, Berkeley, CA, 94720 USA\\ 

$^4$Chemical Sciences Division, Lawrence Berkeley National Laboratory, Berkeley, CA, 94720 USA 
\end{center}

\section{Methods}
\textbf{Charge Transfer Decomposition Analysis.} The ALMO-EDA\cite{horn2016probing} scheme separates the overall intermolecular interaction energy, $\Delta E_\text{INT}$, into contributions from frozen interaction ($\Delta E_\text{FRZ}$), polarization ($\Delta E_\text{POL}$), and charge transfer ($\Delta E_\text{CT}$) from the corresponding intermediate states
\begin{equation}
    \Delta E_\text{INT}=\Delta E_\text{FRZ}+\Delta E_\text{POL}+\Delta E_\text{CT}
\end{equation}
The frozen state is obtained by directly antisymmetrizing the isolated fragment wavefunctions. The polarization state is defined by the relaxation of the molecular orbitals of each fragment in the presence of other fragments and is CT-free.\cite{ge2017simulating,mao2019accurate} In the large basis set limit, orbital overlap between fragments can cause charge transfer contamination in the polarized state. The second generation of ALMO-EDA resolved this by retaining only fragment electric response functions\cite{horn2015polarization,aldossary2025uncoupled}(FERFs), ensuring accurate polarized states even at the complete basis set limit.  Finally, the CT state is achieved from the SCF calculation of the whole complex without any constraint, and it includes the electron flow to and from each molecule in the system due to intermolecular relaxation or mixing of the molecular orbitals.\cite{khaliullin2008analysis, shen2022generalization} The net direction of electron flow is determined by the balance between forward and backward contributions, processes that contribute to stabilizing the system by lowering its energy. 

In the ALMO-EDA framework the amount of CT is defined as 
\begin{equation}
    \Delta Q = \text{Tr}\{\hat{P}_\text{CT}\}-\text{Tr}\{\hat{P}_\text{POL}\hat{P}_\text{CT}\},
\end{equation}
where $\hat{P}_\text{POL}$ and $\hat{P}_\text{CT}$ are the density operators of the POL and CT states. $\Delta Q$ is the electron count in the final state that lies outside the density operator of the initial state, and is identical\cite{shen2024occupied} to the excitation number.\cite{barca2018excitation}  The CT  decomposition analysis of ALMO-EDA breaks this quantity into fragment additive pairs. For instance, the amount of charge transfer from fragment $x$ to fragment $y$ is obtained as
\begin{equation}
    \Delta Q_{x\rightarrow y}=\sum_{ia}\langle\psi^{ya}|\hat{P}^\text{eff}_{vo}|\psi_{xi}\rangle\langle\psi^{xi}|\hat{X}_{ov}|\psi_{ya}\rangle,
\end{equation}
where $\hat{P}^\text{eff}_{vo}$ is an effective density operator, $\hat{X}_{ov}$ is the generator of the unitary transformation connecting $\hat{P}_\text{POL}$ and $\hat{P}_\text{CT}$, and the $\psi$'s are the covariant and contravariant ALMOs. We can rotate the occupied and virtual ALMOs such that $\hat{X}_{ov}$ is diagonal under the rotated orbitals. These special orbitals form a most compact basis for describing the charge transfer, and they are called the complementary occupied-virtual pairs (COVP). Typically, the CT between two fragments in the CT process can be described by one or two dominant COVPs. 

Given the balance between computational cost and accuracy, we selected def2-TZVPD as the optimal basis set for EDA (Details in Supplementary Figure S3). We note on technical grounds the basis set used by Poli and co-workers did not include diffuse functions\cite{poli2020charge}, although diffuse functions are necessary for properly describing electron density delocalization and hydrogen bond interactions, and their absence leads to  misrepresenting the extent of CT in hydrogen-bonded networks\cite{papajak2010efficient}.

Due to the substantial size of the interfacial hexane–water clusters, which contain on average $\sim$400 atoms per cluster, performing EDA at the $\omega$B97X-V/def2-TZVPD level of theory poses a significant computational challenge, particularly for the evaluation of the polarized state. To alleviate the computational cost associated with polarization, we employed two key strategies to enhance efficiency. First, each hexane molecule was assigned as an independent fragment, thereby avoiding the overhead introduced by large fragments. All water molecules were treated collectively as a single fragment. Second, we utilized the recently developed uncoupled-FERFs (uFERFs)\cite{aldossary2025uncoupled}, which provides an enormous speedup through a diagonal approximation of the orbital hessian matrix. The previous benchmark results \cite{aldossary2025uncoupled} demonstrated that the polarization energies of neutral systems are almost unaffected by uFERF approximation.

\textbf{Molecular dynamics simulations }
The system, consisting of 1450 water molecules and 200 hexane molecules, was simulated using the AMOEBA polarizable force field within the OpenMM\cite{eastman2023openmm} platform and all calculations were done in double precision. The dimensions of the simulation box were set to 3.5 × 3.5 × 7.5 nm$^3$. Simulations were kept at a constant temperature of 298.15 K and a constant P$_{zz}$ of 1 atm, where P$_{zz}$ is the component of the pressure tensor perpendicular to the interface (only the 
z-dimension of the periodic simulation box is allowed to change while the x- and y-dimensions are kept fixed).  The simulations used the MTSLangevinIntegrator for efficient temperature control and integration. Non-bonded interactions were handled using the Particle Mesh Ewald (PME) method, with a non-bonded cutoff of 1.2 nm and an Ewald error tolerance of 0.0005. The system was equilibrated for 1 ns, with a time step of 0.5 fs. Following equilibration, a 20 ns production run was conducted to investigate the structural properties of the water-hexane interface. Snapshots were saved every 100 ps, yielding a total of 200 frames for EDA analysis.

\textbf{Interfacial hexane-water clusters selection} Based on the analysis of the radial distribution function (RDF) and the two-dimensional probability density distribution, a C–O distance cutoff of 4.0 ~\AA \ and an angular criterion of $\angle(\text{C–H$\ldots$O}) \geq 100^\circ$ were defined to identify hydrogen bonds between water and hexane. Within this distance range, the angle distribution remains relatively constrained, with a pronounced preference for angles exceeding 100°, supporting the validity of the chosen threshold. In addition to direct water–hexane interactions, interfacial water molecules form an extended hydrogen-bond network surrounding the hexane molecules. To characterize these networks, hydrogen bonding among water molecules was identified using geometric criteria: an O-O distance cutoff of 3.0 ~\AA \ and an angular criterion of $\angle(\text{O–H$\ldots$O}) \geq 120^\circ$. Within this expanded network, interfacial water molecules engage in additional hydrogen bonds with neighboring water molecules, further stabilizing the interfacial configuration through cooperative interactions.
To ensure computational feasibility, the interfacial area was reduced to 4 nm$^2$. A hexane molecule was considered interfacial if at least half of its carbon atoms were located within this defined region.

\section{Additional Methods}

\begin{figure}[H]
    \centering
    \includegraphics[width=0.8\textwidth]{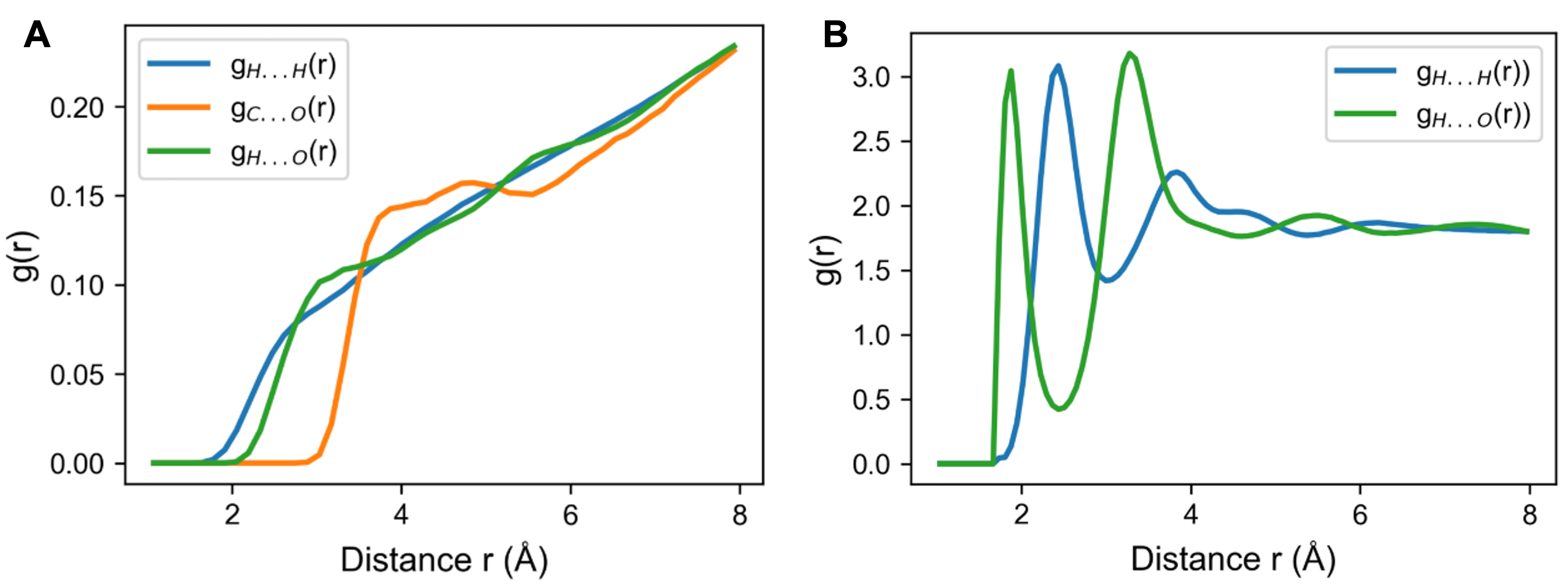}
    \caption{Radial distribution functions (RDFs) of (\textbf{A}) hexane-water and (\textbf{B}) water-water. Each function g$_{\text{A$\ldots$B}}$(r) describes the probability of finding atom A of hexane(left) or water(right) at a distance r from atom B in water.}
    \label{fig:gr}
\end{figure}

\noindent
\textbf{RDF for the water oil interface.} Supplementary Figure \ref{fig:gr}A presents the radial distribution functions (RDFs) of key atomic pairs at the water–hexane interface. In the water–hexane RDF , the shortest intermolecular distance corresponds to H···H (~2.0 ~\text{\AA}) rather than O···H, suggesting that hydrogen atoms from water are preferentially interacting with hexane molecules. In contrast, Supplementary Figure \ref{fig:gr}B, the water–water RDF shows that due to strong hydrogen bond formation, the O···H peak is located at $\sim$1.9 ~\text{\AA}. Compared to water dimers, where the O···H hydrogen bonds form at shorter distances and dominate the interaction, the water–hexane interface exhibits an   O···H peak at ~3.0 ~\text{\AA}, indicating weaker hydrogen bonding.

\vspace{3mm}

\begin{figure}[H]
    \centering
    \includegraphics[width=0.8\textwidth]{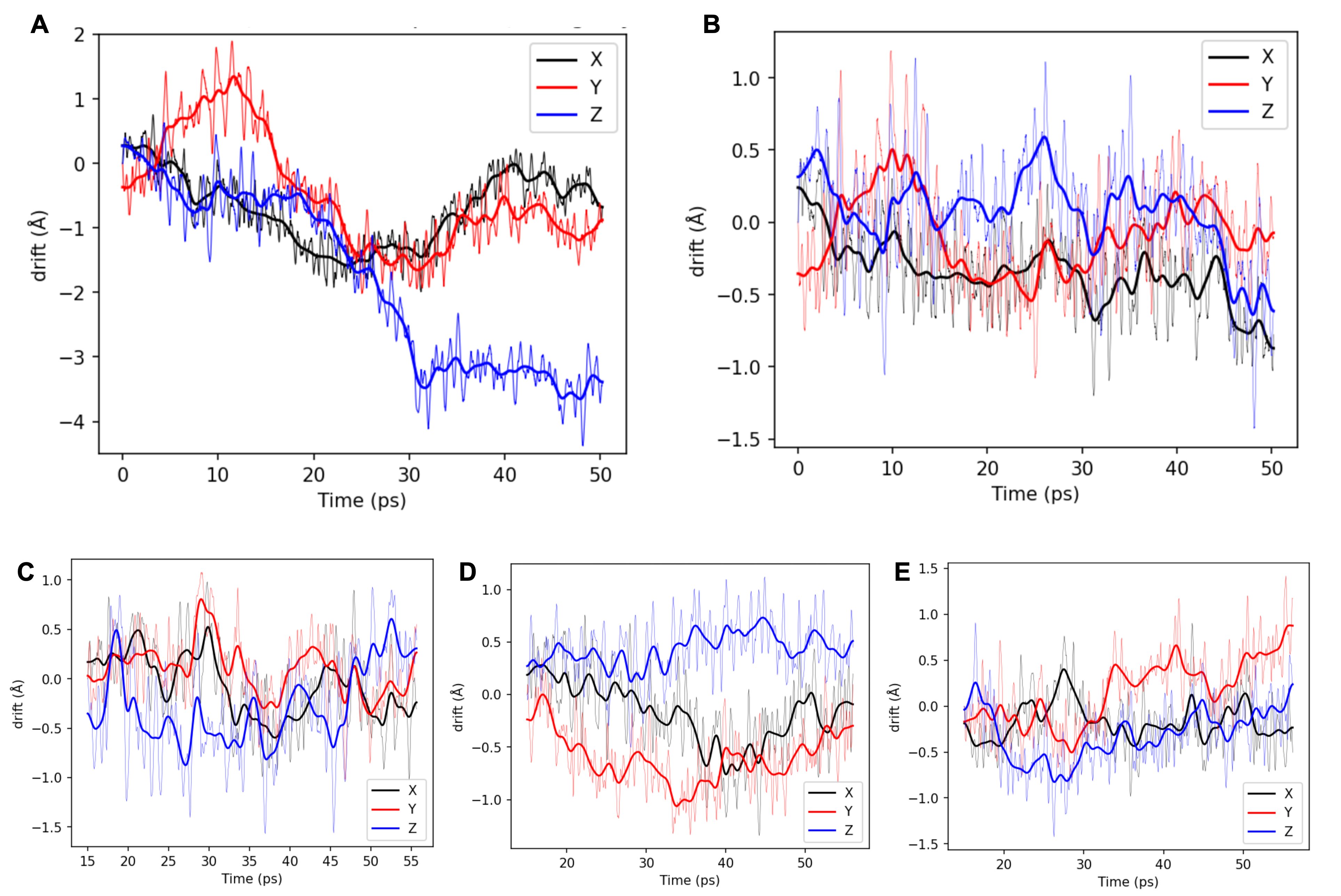}
    \caption{(\textbf{A}) The drift of the center of mass (COM) of carbon atoms in neopentane, which closely matches the results presented in Fig. 3D of the paper\cite{Pullanchery2024}. (\textbf{B}) The net drift of neopentane carbons after subtracting the center-of-mass motion of the entire simulation box.\rev{ (\textbf{C, D and E}) The net drift of neopentane carbons from three independent simulations.} Dashed lines represent raw data; solid thick lines show the running averages.}
    \label{fig:neopentane-drift}
\end{figure}

\noindent
\textbf{Neopentane drift analysis. } To further examine the drift of neopentane under a 0.1 V/~\text{\AA} electric field, we analyzed the trajectory data corresponding to Fig. 3D from the original study (available at Zenodo: https://doi.org/10.5281/zenodo.11532589). The computed drift of the center of mass (COM) of neopentane's carbon atoms closely matched the reported data (Figure \ref{fig:neopentane-drift}A). However, visualization of the trajectory in VMD revealed an overall drift of the entire system. Further analysis indicated that neopentane's movement was highly correlated with the drift of surrounding water molecules, both collectively and for individual water molecules initially within 3.5 ~\text{\AA} of neopentane. After subtracting the center-of-mass motion of the simulation box, the net drift of neopentane carbons along the x, y, and z directions was found to be -0.78 ~\text{\AA}, -0.11 ~\text{\AA}, and -0.60 ~\text{\AA}, respectively, over the 50 ps timescale as seen in Figure \ref{fig:neopentane-drift}B. These results suggest that neopentane does not exhibit a significant independent drift along the z-axis.

We also performed three independent simulations of a system containing one neopentane molecule and 235 water molecules under a 0.1 V/Å electric field oriented along the z-axis. Following an equilibration period of 15 ps, about 30 ps production trajectories were collected. Figure \ref{fig:neopentane-drift}C, D and E summarize the drift of neopentane carbons from these simulations. The net drift calculated from our independent simulations are less than 1 Å and do not exhibit any significant specificity along the z-axis

\vspace{3mm}

\noindent
\textbf{Nonlinear Poisson-Boltzmann Equation.} The potential distribution \( \psi(r) \) can be described by the Poisson equation:
\begin{equation}
\nabla^2 \psi = \frac{\partial^2 \psi}{\partial x^2} + \frac{\partial^2 \psi}{\partial y^2} + \frac{\partial^2 \psi}{\partial z^2} = -\frac{\rho_e}{\varepsilon_r \varepsilon_0},
\end{equation}
where \( \rho_e \) is the charge density, \( \varepsilon_r \) is the relative permittivity, and \( \varepsilon_0 \) is the vacuum permittivity. According to the Boltzmann distribution, the local ion density \( c_i \) is defined as:
\begin{equation}
c_i = c_i^0 \cdot e^{-W_i / k_B T},
\end{equation}
where \( c_i^0 \) is the reference ion density, \( W_i \) is the potential energy of the ion, \( k_B \) is the Boltzmann constant, and \( T \) is the absolute temperature.

The Poisson-Boltzmann equation combines the relationship between position and potential as:
\begin{equation}
\nabla^2 \psi = \frac{c_0 e}{\varepsilon_r \varepsilon_0} \cdot \left( e^{e \psi(x, y, z) / k_B T} - e^{-e \psi(x, y, z) / k_B T} \right),
\end{equation}
where \( c_0 \) is the bulk ion concentration and \( e \) is the elementary charge. This equation demonstrates the coupling between the electrostatic potential and the ionic distribution, forming the basis for describing ion behavior in electrolyte solutions.

\vspace{3mm}

\noindent
\textbf{Surface Charge Density and surface potential definition. } The surface charge density \( \sigma \) is determined by the boundary condition, where the surface charge plus the charge of the ions in the entire double layer must equal zero:
\begin{equation}
\sigma = - \int_R^\infty \rho_e \, dx,
\end{equation}
where \( \rho_e \) is the local charge density. Using the Poisson equation, this can be rewritten as:
\begin{equation}
\sigma = \varepsilon_0 \varepsilon_r \int_R^\infty \frac{d^2 \psi}{dx^2} \, dx = -\varepsilon_0  \varepsilon_r \frac{d \psi}{dx} \bigg|_{x=R},
\end{equation}
where \( \varepsilon_0 \) is the vacuum permittivity, and \( \psi \) is the electrostatic potential.

The relationship between the surface potential \( \psi_0 \) and the surface charge density \( \sigma_0 \) can be expressed as:
\begin{equation}
\sigma_0 = Z \sqrt{2 \varepsilon_r \varepsilon_0 k_B T \rho_\text{bulk}} 
\left[
2 \sinh\left( \frac{e \psi_0}{2 k_B T} \right)
+ \frac{4}{R \left( \frac{\varepsilon_0 \varepsilon_r k_B T}{2 e^2 Z^2 \rho_\text{bulk}} \right)^{-1/2}}
\tanh\left( \frac{e \psi_0}{4 k_B T} \right)
\right],
\end{equation}
where:
- \( Z \): Ion valence,
- \( \varepsilon_r \): Relative permittivity,
- \( k_B \): Boltzmann constant,
- \( T \): Absolute temperature,
- \( \rho_\text{bulk} \): Bulk ion density,
- \( R \): Distance from the surface,
- \( e \): Elementary charge. This equation highlights the dependence of the surface charge density on the surface potential, ion concentration, and other system parameters. We note that previous experiments have reported average surface droplet radii of 100 nm\cite{pullanchery2021charge,Shi2025} to 230 nm\cite{pullanchery2020stability,Pullanchery2024}, with larger radii diminishing estimated surface potential magnitudes; here we use a radius of 125 nm as reported most recently\cite{Pullanchery2024,Shi2025}. 

\vspace{3mm}

\noindent
\textbf{Origin data for small and large interface clusters.} The CT behavior for the water dimer and water-hexane small clusters in Figure 1 in the main text are given in Supplementary Tables \ref{tab:CT_analysis} and \ref{tab:eda_analysis}.
The CT behavior at the water-hexane interface was evaluated using EDA-COVP analysis on 200 clusters, which were divided into two independent groups of 100 clusters each. As shown in Supplementary Table \ref{tab:interface-CT}, the forward, backward, net, and total CT values, expressed in me$^-$/nm$^2$, are highly consistent between the two sets. For instance, the average forward CT values are 2.15 and 2.07 for Set 1 and 2, respectively, while the net CT values are 0.68 and 0.63 me$^-$/nm$^2$. Similarly close agreements are observed for the backward and total CT metrics. This consistency not only validates our selection criteria and computational methodology but also reinforces the robustness of our simulation approach in capturing the interfacial charge transfer properties at the water–oil interface.
\vspace{3mm}

\begin{figure}[h]
    \centering
    \includegraphics[width=0.8\textwidth]{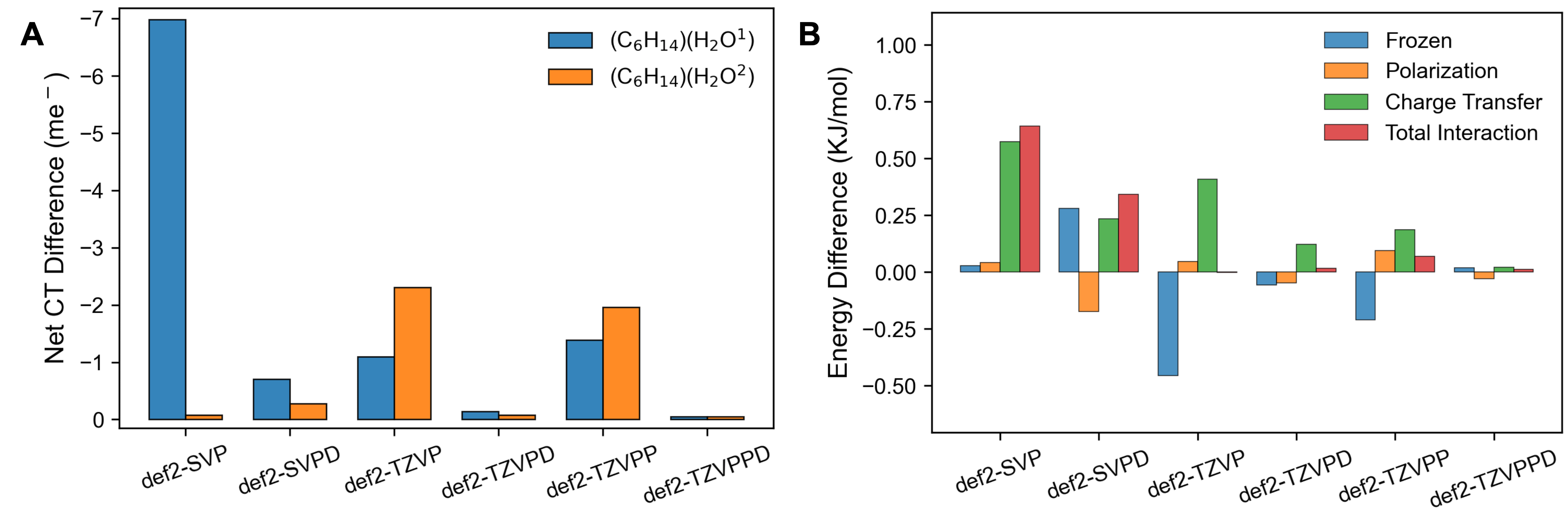}
    \caption{(\textbf{A}) Net CT differences for structures C$^1$ and C$^2$ relative to def2-QZVPPD. (\textbf{B}) Energy decomposition differences (Frozen, Polarization, CT, and Total) relative to def2-QZVPPD. }
    \label{fig:basis-set-difference}
\end{figure}

\noindent
\textbf{Basis Set Dependence on EDA components and CT value. } To evaluate the effect of basis set choice on both the computed net CT and  EDA (Frozen, Polarization, CT, and Total Interaction Energy) components, we analyzed structures C$^1$ and C$^2$ in the main text using various basis sets in Figure \ref{fig:basis-set-difference}; the results are compared to those obtained with the def2-QZVPPD basis, which was taken as the high quality reference. The data reveals that the def2-TZVPD basis set provides the best agreement with the reference basis for both properties. For net CT, the deviations from def2-QZVPPD are $-0.13$ for C$^1$ and $-0.08$ for C$^2$, while def2-SVP, def2-TZVP, and def2-TZVPP exhibit larger discrepancies. This confirms that the CT component is highly sensitive to diffuse functions, and that smaller basis sets introduce significant inaccuracies. Given the balance between computational cost and accuracy, we selected def2-TZVPD as the optimal basis set for all calculations reported in the main paper.

\newpage
\section{Supplementary Figures}

\section{Supplementary Tables}

\begin{table}[H]
\centering
\begin{tabular}{lccccc}
\hline
\textbf{Complex}  & \textbf{Total CT} & \textbf{Backward CT} & \textbf{Forward CT}& \textbf{net CT} \\
\hline
(H$_2$O)(H$_2$O)                 &2.62  & 0.14 & 2.81 &2.67\\
(C$_6$H$_{14}$)(H$_2$O)               & 0.65 & 0.32 & 0.42 & 0.10 \\
(C$_6$H$_{14}$)(H$_2$O$^1$ $\ldots$ H$_2$O$^2$)      & 1.29 & 0.71 & 0.82 & 0.11 \\
(C$_6$H$_{14}$)(H$_2$O$^1$)      & 0.56 & 0.11 & 0.47 & 0.36 \\
(C$_6$H$_{14}$)(H$_2$O$^2$)      & 0.72 & 0.55 & 0.27 & -0.28 \\
\hline
\end{tabular}
\caption{Charge transfer (CT) analysis for water hexane complexes. Forward CT refers to H$_2$O $\to$ C$_6$H$_{14}$, and Backward CT refers to C$_6$H$_{14}$ $\to$ H$_2$O, in units of me$^-$, and using $\omega$B97X-V/def2-TZVPD.}
\label{tab:CT_analysis}
\end{table}

\begin{table}[H]
\centering
\begin{tabular}{lcccc}
\hline
\textbf{Complex}  & \textbf{Froz} & \textbf{Pol} & \textbf{CT} & \textbf{Total} \\
\hline
(H$_2$O)(H$_2$O)             &-9.3  & -4.5 & -6.7 & -20.5 \\
(C$_6$H$_{14}$)(H$_2$O) & -3.5 & -1.6 & -1.4 & -6.5 \\
(C$_6$H$_{14}$)(H$_2$O$^1$ $\ldots$ H$_2$O$^2$) & -6.2 & -4.6 & -2.8 & -13.7 \\
(C$_6$H$_{14}$)(H$_2$O$^1$)  & -3.8 & -1.3 & -0.9 & -6.0\\
(C$_6$H$_{14}$)(H$_2$O$^2$) & -2.1 & -1.7 & -1.7 & -5.4\\
\hline
\end{tabular}
\caption{ALMO-EDA for water dimer and water hexane complexes (in KJ/mol) , basis and functional $\omega$B97X-V/def2-TZVPD.}
\label{tab:eda_analysis}
\end{table}

\begin{table}[H]
\centering
\begin{tabular}{lccc}
\hline
\textbf{CT Type} & \textbf{Average} & \textbf{Max} & \textbf{Min} \\
\hline
\multicolumn{4}{c}{\textbf{First Set (100 Clusters)}} \\
Forward  & 2.15 & 3.58 & 1.07 \\
Backward & -1.48 & -0.61 & -2.48 \\
Net      & 0.68 & 2.03 & -0.07 \\
\hline
\multicolumn{4}{c}{\textbf{Second Set (100 Clusters)}} \\
Forward  & 2.07 & 3.14 & 1.16 \\
Backward & -1.44 & -0.76 & -2.72 \\
Net      & 0.63 & 1.34 & -0.01 \\
\hline
\end{tabular}
\caption{Charge transfer analysis (me$^-$/nm$^2$) for two independent sets of 100 clusters. Description of the two clusters are found in Figure 3A in the main text.}
\label{tab:interface-CT}
\end{table}

\begin{table}[H]
\centering
\begin{tabular}{lccc}
\hline
\textbf{CT Type} & \textbf{Average} & \textbf{Max} & \textbf{Min} \\
\hline
Forward  & 2.09 & 3.22 & 1.21 \\
Backward & -1.48 & -0.77 & -2.37 \\
Net      & 0.61 & 1.44 & -0.05 \\
\hline
\end{tabular}
\caption{Charge transfer analysis (me$^-$/nm$^2$) for the sets of 50 clusters. Description of the clusters are found in Figure 3B in the main text.}
\label{tab:interface-CT}
\end{table}

\bibliography{reference}
\bibliographystyle{naturemag}